\documentclass[twocolumn,showpacs,preprintnumbers,amsmath,amssymb]{revtex4}

\usepackage{dcolumn}%
\usepackage{epsfig}

\begin{document}
\title{\bf Extremal Optimization at the Phase Transition of the
3-Coloring Problem}

\author{Stefan Boettcher} \email{sboettc@emory.edu}
\affiliation{Physics Department, Emory University, Atlanta, Georgia
30322, USA} 
\author{Allon G.\ Percus} \email{percus@ipam.ucla.edu}
\affiliation{Computer \& Computational Sciences Division, Los Alamos
National Laboratory, Los Alamos, NM 87545, USA \\
and UCLA Institute for Pure and Applied Mathematics, 
Los Angeles, CA 90049, USA}
\date{\today}
\begin{abstract} 
We investigate the phase transition of the 3-coloring problem on
random graphs, using the extremal optimization heuristic.  3-coloring
is among the hardest combinatorial optimization problems and is
closely related to a 3-state anti-ferromagnetic Potts model. Like many
other such optimization problems, it has been shown to exhibit a phase
transition in its ground state behavior under variation of a system
parameter: the graph's mean vertex degree.  This phase transition is
often associated with the instances of highest complexity.  We use
extremal optimization to measure the ground state cost and the
``backbone'', an order parameter related to ground state overlap,
averaged over a large number of instances near the transition for
random graphs of size $n$ up to 512. For graphs up to this size,
benchmarks show that extremal optimization reaches ground states and
explores a sufficient number of them to give the correct backbone
value after about $O(n^{3.5})$ update steps. Finite size scaling
gives a critical mean degree value $\alpha_{\rm
c}=4.703(28)$. Furthermore, the exploration of the degenerate ground
states indicates that the backbone order parameter, measuring the
constrainedness of the problem, exhibits a first-order phase
transition.
\end{abstract} 
\maketitle

\section{Introduction}
\label{intro}
The most challenging instances of computational problems are often
found near a critical threshold in the problem's parameter
space~\cite{HH}, where certain characteristics of the problem change
dramatically. One such problem, already discussed in
Ref.~\cite{MPWZ,Cheese,Hogg,Culberson}, is the 3-coloring problem.
Consider a random graph~\cite{Bollobas} having $n$ vertices and $m$
edges placed randomly among all possible pairs of vertices. The number
of edges emanating from each vertex is then Poisson-distributed around
a mean degree $\alpha=2m/n$. To 3-color the graph, we need to assign
one of three colors to each vertex so as to minimize the number of
``monochromatic'' edges, i.e., those connecting vertices of the same
color. In particular, we may want to decide whether it is possible to
make an assignment without any monochromatic edges using, for
instance, a backtracking assignment procedure. Typically, if the mean
degree $\alpha$ is low (for example, when each vertex most likely has
fewer than 3 neighbors), one quickly finds a perfect coloring. If the
mean degree is high, one soon determines that monochromatic
edges are unavoidable after fixing just a small number of vertices. At
an intermediate mean degree value, however, some graphs are perfectly
colorable while others are not. In that case, for each instance one
must inspect many almost-complete colorings, most of which fail only
when trying to assign a few final vertices, before colorability can be
decided~\cite{Hogg,C+M}.  For increasing $n$, the regime of mean
degree values $\alpha$ for which the decision problem is hard becomes
narrowly focused, while the computational complexity of the
backtracking algorithm within this regime grows faster than any
power of $n$, signs of the impending singularity associated with a
phase transition.

Such findings have spawned considerable interest among computer
scientists and statistical physicists alike. On one hand, there appear
to be close links to the properties of spin glass systems~\cite{MPV}.
Using replica symmetry breaking, it was recently
argued~\cite{MPWZ,BMPWZ} that 3-coloring undergoes a colorability
transition at $\alpha_{\rm crit}=4.69$, heralded by the spontaneous
emergence at $\alpha=3.35$ of a sizable 3-core~\cite{PSW} that becomes
over-constrained at the transition. This analysis shows, furthermore,
that the hardest instances to decide are located between a clustering
transition at $\alpha\approx4.42$ and $\alpha_{\rm crit}$.  On the
other hand, attempts have been made to relate the nature of the phase
transition to complexity classifications developed by computer
scientists for combinatorial problems~\cite{G+J}: it has been
suggested that NP-complete problems, which are hard to solve, display
a first-order phase transition while easier problems lead to a
second-order transition~\cite{Monasson}.  Such a relation, while
intriguing and suggestive, is bound to be questionable in light of the
fact that these phase transitions are based on a notion of
average-case complexity (for instance, 3-coloring averaged over the
ensemble of random graphs), distinct from the notion of worst-case
complexity used by computer scientists to define
NP-completeness~\cite{Cook} and other such complexity theoretic
categories. In fact, it currently appears that a first-order
transition is merely an indicator for the complexity of certain types
of algorithms: local searches.  A model problem with a discontinuous
transition, $K$-XORSAT~\cite{Ricci,Leone}, can be solved by a fast
global algorithm yet it is extremely hard for local search or
backtracking assignment algorithms.

In this paper we consider the 3-coloring problem mentioned above. The
problem is among the hardest combinatorial optimization problems,
making it difficult to study the asymptotic properties of its phase
transition with exact methods. It is also of considerable interest in
its own right as a model for many practical optimization
problems~\cite{G+J}, and it is of some physical relevance due to its
close relation to Potts anti-ferromagnets~\cite{colorcites}. Some
aspects of the 3-coloring phase transition have previously been
explored~\cite{Cheese,Hogg,Culberson}. In particular, Culberson and
Gent~\cite{Culberson} have studied the phase transition for random
graphs with {\em exact} methods by ``growing'' random graphs of size
$n\leq225$, sequentially adding random edges to an existing graph to
increase $\alpha$, and checking along the way whether the graph is
still colorable.  Once a graph becomes uncolorable, it is discarded
from the list of growing graphs, so the set of graphs becomes
increasingly less representative of the ensemble when passing through
the transition. In the process, these authors have evaluated the
constrainedness of the variables in the graph, studying in detail many
aspects of the approach to the transition. However, the main quantity
they measure, called the ``spine''~\cite{Chayes}, is in general an
upper bound on the order parameter we consider here, since
contributions from uncolorable graphs are neglected.

We investigate the properties near the phase transition by applying an
optimization heuristic called {\em extremal optimization}
(EO)~\cite{BoPe1,eo_prl}. EO was recently introduced as a
general-purpose optimization method based on the dynamics of driven,
dissipative systems~\cite{PPSN}. Our study shows that EO is capable of
determining many ground-state properties efficiently, even at the
phase transition ~\cite{eo_perc}.  EO performs a local neighborhood
search that does not get stuck in local minima but proceeds to explore
near-optimal configurations broadly. Hence, it is particularly well
suited to measure global properties of the configuration space.
Here, we use it to estimate the ``backbone'', an overlap property between
the highly degenerate ground state configurations that provides a more
convenient order parameter than measuring mutual overlaps of all
ground states~\cite{Monasson}. While EO is not exact, benchmark
comparisons with exactly-solved, large instances justify our
confidence in its results. Our biggest errors originate from the lack
of statistics at large $n$.

Our results indicate that the transition in the backbone size is of
first order, though with only a small discontinuity. In fact, the
discontinuity does not arise uniformly for all graphs in the ensemble,
but is due to a fraction of instances that have a strong backbone of a
characteristic size while the rest have hardly any backbone at all.

Using the procedure of Ref.~\cite{Somen} to control the quality of
finite size scaling for the ground state cost function, we estimate
the location of the transition as $\alpha_{\rm crit}\approx4.703(28)$,
where the numbers in parentheses denote the statistical error bar in
the final digits.  This is consistent with the presumably correct
value of $\alpha=4.69$ given by replica symmetry breaking
methods~\cite{MPWZ} (see also Refs.~\cite{Hogg,Cheese} for earlier
estimates).  We measure the size of the scaling window as $n^{-1/\nu}$
with $\nu=1.43(6)$, close to the value of 1.5 estimated for
3-SAT~\cite{Selman}, although it may be that trivial
$n^{-1/2}$-fluctuations from the variables not belonging to the
3-core~\cite{PSW} dominate at much larger $n$ than considered
here~\cite{Wilson}.

In the following section, we introduce the problem of 3-coloring in
more detail and discuss the relevant observables we measure in order
to analyze the phase transition. In Sec.~\ref{EOalgo} we discuss our
EO implementation and its properties. In Sec.~\ref{numerics} we
present the results of our measurements, and we conclude with
Sec.~\ref{conclusion}.

\section{3-Coloring of random graphs}
\label{3col}
A random graph~\cite{Bollobas} is constructed from a set of $n$
vertices by assigning an edge to $m=\alpha n/2$ of the ${n \choose 2}$
pairs of vertices with equal probability, so that $\alpha$ is the
average vertex degree.  Here, we will only consider the regime of
``sparse'' random graphs where $m=O(n)$ and $\alpha=O(1)$.  The goal
of graph coloring is to label each vertex with a different color so as
to avoid monochromatic edges.

Three different versions of the coloring problem are of interest.
First, there is the classic problem of determining the ``chromatic
number'' for a given graph, i.e., the minimum number of colors needed
to color the graph while avoiding monochromatic edges. It is very
difficult to devise a heuristic for this problem~\cite{JohnsonCOL}. In
the other two versions, we are given a fixed number $K$ of colors to
select from. The decision problem, $K$-COL, addresses the question of
whether a given graph is colorable or not. Finally, the optimization
problem, MAX-$K$-COL, tries to minimize the number of monochromatic
edges (or equivalently, maximize the number of non-monochromatic
edges, hence its name).  Clearly, if we define the number of
monochromatic edges as the ``cost'' or ``energy'' of a color
assignment, determining whether the minimal cost is zero or non-zero
corresponds to solving the decision problem $K$-COL, so finding the
actual cost of the ground state is always at least as hard. Much of
the discussion regarding the complexity near phase transitions in the
computer science literature is focused on the decision
problem~\cite{Cheese,Culberson}. From a physics perspective, it seems
more intuitive to examine the behavior of the ground states as one
passes the transition. Accordingly, we will focus on the MAX-$K$-COL
problem in this paper.

All these versions of coloring are NP-hard~\cite{G+J}, and thus
computationally hard in the worst case.  To determine exact answers
would almost certainly require a computational time growing faster
than any power of $n$. Thus, extracting results about asymptotic
properties of the problems is a daunting task, calling for the use of
accurate heuristic methods, as discussed in the following section.

The control parameter describing our ensemble of instances is the
average vertex degree $\alpha$ of the random graphs.  Constructing an
appropriate order parameter to classify the transition is less
obvious.  The analogy to spin-glass theory~\cite{MPV,F+H} suggests the
following reasoning. In a homogeneous medium possessing a single pure
equilibrium state, the magnetization provides the conjugate field to
analyze the ferromagnetic transition. For our 3-coloring problem, the
disorder induced by the random graphs leads to a decomposition into
many coexisting but unrelated pure states with a distribution of
magnetizations. Since the colors correspond to the spin orientations
in the related Potts model, we need in principle to determine, for
each graph, the overlap between all pairs of ground state
colorings. Finally, this distribution has to be averaged over the
ensemble. To simplify the task, one can instead extract directly the
``backbone'', which is the set of variables that take on the same
state in {\em all} possible ground state colorings of a given
instance~\cite{Monasson}. But even determining the backbone is a
formidable undertaking: it requires not only finding a lowest cost
coloring but sampling a substantial number of those colorings for each
graph, since the ground state entropy is extensive.

Another level of difficulty arises due to the invariance of the ground
states under a global color permutation. Thus, in the set of all
ground states, each vertex can take on any color and the backbone as
defined above is empty. To avoid this triviality, one may redefine the
backbone in the following way. Instead of considering individual
vertices, consider all {\em pairs} of vertices that are not connected
by an edge~\cite{Culberson}. Define the pair to be part of the {\em
frozen} backbone if its vertices are of like color (monochromatic) in
all ground state colorings, so that the presence of an edge there
would necessarily incur a cost.  Define the pair to be part of the
{\em free} backbone if its vertices are of unlike color
(non-monochromatic) in all ground state colorings, so that the
presence of an edge there would not incur any cost.  Since the
fraction of pairs that belong to the frozen backbone measures the
constrainedness of an instance, it is the relevant order parameter. We
have also sampled the free backbone. As shown in Sec.~\ref{numerics},
both seem to exhibit a first-order transition, though the jump for the
frozen backbone is small.

By definition, the location of the transition is determined through a
(second-order) singularity in the cost function $C$: the cost is
asymptotically vanishing below the transition, it is continuous at the
transition, and above it is always non-zero, We have therefore
measured the ground state cost, averaged over many instances, for a
range of mean degree values $\alpha$ and sizes $n$.

\section{Extremal Optimization}
\label{EOalgo}
To investigate the phase transition in 3-COL, we employ the extremal
optimization heuristic (EO)~\cite{BoPe1}. The use of a heuristic
method, while only approximate, allows us to measure observables for
much larger system sizes $n$ and with better statistics then would be
accessible with exact methods. We will argue below that we can obtain
optimal results with sufficient probability that even systematic
errors in the exploration of ground states will be small compared to
the statistical sampling error.

Our EO implementation is as follows. Assume we are given a
graph with a (however imperfect) initial assignment of colors to its
vertices. Each vertex $i$ has $\alpha_i$ edges to neighboring
vertices, of which $0\leq g_i\leq\alpha_i$ are ``good''
edges, i.e., to neighbors of a different color (not monochromatic).
We define for each vertex a ``fitness''
\begin{eqnarray}
\lambda_i=\frac{g_i}{\alpha_i}\in[0,1],
\end{eqnarray}
and determine a permutation $\Pi$ (not necessarily unique) over the vertices
such that
\begin{eqnarray}
\lambda_{\Pi(1)}\leq\lambda_{\Pi(2)}\leq\ldots\leq\lambda_{\Pi(n)}.
\label{lambdaeq}
\end{eqnarray}
At each update step, EO draws a number $k$ from a distribution
\begin{eqnarray}
P(k)\sim k^{-\tau}\quad(1\leq k\leq n)
\label{taueq}
\end{eqnarray}
with a bias toward small numbers. A vertex $i$ is selected from the
ordered list in Eq.~(\ref{lambdaeq}) according to its ``rank'' $k$,
i.e., $i=\Pi(k)$. Vertex $i$ is updated {\em unconditionally}, i.e.,
it always receives a new color, selected at random from one of the
other colors. As a consequence, vertex $i$ and all its neighbors
change their fitnesses $\lambda$ and a new ranking $\Pi$ will have to
be established. Then, the update process starts over with selecting a
new rank $k$, and so on until some termination condition is
reached. Along the way, EO keeps track of the configuration with the
best coloring it has visited so far, meaning the one that minimizes
the number of monochromatic edges, $C=\sum_i(\alpha_i-g_i)/2$.

Previous studies have found that EO obtains near-optimal solutions for
a variety of hard optimization problems~\cite{PPSN} for a carefully
selected value of $\tau$~\cite{BoPe2,eo_jam,Dall}. For 3-COL, initial
trials have determined that best results are obtained for the system
sizes $n=32,64,\ldots,512$ at a (fixed) value of $\tau\approx2.2$. This
rather large value of $\tau$~\cite{eo_jam} helps explore many low-cost
configurations efficiently; if we merely wanted to determine low-cost
solutions, larger values of $n$ could have been reached
more efficiently at a smaller value of $\tau$.

It should be noted that our definition of fitness does not follow the
generic choice $\lambda_i=g_i/2$ that would give a total configuration
cost of $C={\rm const.}-\sum_i\lambda_i$.  While this formulation
sounds appealing, and does produce results of the same quality, our
choice above produces those same results somewhat faster; there
appears to be some advantage to treating all vertices, whose
individual degrees $\alpha_i$ are Poisson-distributed around the mean
$\alpha$, on an equal footing. Furthermore, our implementation limits
itself to partially sorting the fitnesses on a balanced
heap~\cite{BoPe1}, rather than ranking them perfectly as in
Eq.~(\ref{lambdaeq}). In this way, the computational cost is reduced
by a factor of $n$ while performance is only minimally
affected~\cite{BoPe1}.

\subsection{Measuring the backbone}
\label{backbone}
The backbone, described in Sec.~\ref{3col}, is a collective property
of degenerate ground states for a given graph.  Thus, in this study we
are interested in determining not only the cost $C$ of the ground
state, but also a good sampling of {\it all} possible ground state
configurations. Local search with EO is ideally suited to probe for
properties that are broadly distributed over the configuration space,
since for small enough $\tau$ it does not get trapped in restricted
regions. Even after EO has found a locally minimal cost configuration,
it proceeds to explore the configuration space widely to find new
configurations of the same or lower cost, as long as the process is
run.

Against these advantages, one must recognize that EO is merely a
heuristic approximation to a problem of exponential complexity. Thus,
to safeguard the accuracy of our measurements, we devised the
following adaptive procedure. For each graph, starting from random
initial colorings, EO was run for $n^3$ update steps, using a minimum
of 5 different restarts. For the lowest cost seen so far, EO keeps a
buffer of up to $n/4$ most recently visited configurations with that
cost. If it finds another configuration with the same cost, it quickly
determines whether it is already in the buffer. If not, EO adds it on
top of the buffer (possibly ``forgetting'' an older configuration at
the bottom of the buffer). Thus, EO does not keep a memory of all
minimal cost configurations seen so far, which for ground states can
have degeneracies of $>10^6$ even at $n=64$. Instead of
enumerating all ground states exhaustively, we proceed as follows.
When EO finds a new, lowest cost configuration, it assumes initially
that all pairs of equally colored vertices are part of the frozen
backbone and all other pairs are part of the free backbone. If another
configuration of the same cost is found and it is not already in the
buffer, EO checks all of the pairs in it. If a pair has always been
frozen (free) before and is so now, it remains part of the frozen
(free) backbone. If a pair was always frozen (free) before and it is
free (frozen) in this configuration, it is eliminated from both
backbones. If a pair has already been eliminated previously, no action
is taken. In this way, certain ground-state configurations may
be missed or tested many times over, without affecting the backbones
significantly.

Eventually, even if new and unrecognized configurations of the lowest
cost are found, no further changes to either backbone are likely to
occur. This fact motivates our adaptive stopping criterion for EO.
Assume the current backbone was last modified in the $r_0$th
restart. Then, for this graph EO restarts for a total of at least
$r=r_0+max\{r_0,5\}$ times, terminating only when there has been no
updates to the backbone over the previous $max\{r_0,5\}$ restarts. Of
course, every time a new, lower-cost configuration is found, the
buffer and backbone arrays are reset. Ultimately, this procedure leads
to adaptive runtimes that depend on the peculiarities of each
graph. The idea is that if the lowest cost state is found in the first
start and the backbone does not change over 5 more restarts, one
assumes that no further changes to it will ever be found by
EO. However, if EO keeps updating the backbone through, say, the 20th
restart, one had better continue for 20 more restarts to be confident
of convergence.  The typical number of restarts was about 10,
while for a few larger graphs, more than 50 restarts were required.

A majority of our computational time is spent merely confirming that
the backbones have converged, since during the final 
$max\{r_0,5\}$ restarts nothing new is found.  Nevertheless, EO still
saves vast amounts of computer time and memory in comparison with
exact enumeration techniques.  The trade-off lies in the risk of
missing some lowest cost configurations, as well as in the risk of
never finding the true ground state to begin with. To get an estimate
of the systematic error resulting from these uncontrollable risks, we
have benchmarked our EO implementation against a number of different
exact results.

First, we used a set of 700 explicitly 3-colorable graphs over 7
different sizes, $n=75,100,\ldots,225$ (100 graphs per value of $n$)
at $\alpha=4.7$, kindly provided by J.~Culberson, for which exact
spine values~\cite{Chayes} were found as described in
Ref.~\cite{Culberson}.  For colorable graphs such as these, the spine
is identical to the backbone. Our EO implementation correctly
determined the 3-colorability of all but one graph, and reproduced
nearly all backbones exactly, regardless of size $n$. For all graph
sizes, EO failed to locate colorable configurations on at most 5
graphs out of 100, and in those cases overestimated either backbone
fraction by less than 4\%. Only at $n=225$ did EO miss the
colorability of a single graph to find $C=1$ instead, thereby
underestimating both backbones.

In a different benchmark, containing colorable as well as
uncolorable graphs, we generated 440 random graphs over 4 different
sizes, $n=32,64,128,256$ and 11 different mean degree values
$\alpha=4.0,4.1,\ldots,5.0$ (10 graphs per value of $n$ and $\alpha$).
We found the exact minimum cost and exact fraction of pairs belonging
to the backbone for these graphs, by removing edges until an exact
branch-and-bound code due to M.A.~Trick~\cite{Trick}
determined 3-colorability.  For example, finding that a graph had a
ground state cost of $C=2$ involved considering all possible 2-edge
removals until a remainder graph was found to be 3-colorable.  We then
added edges to vertex pairs in this remainder graph, checking whether
the graph stayed colorable: if so, that pair was eliminated from the
frozen backbone.  Likewise, we merged vertex pairs, checking whether
the graph stayed colorable: if so, that pair was eliminated from the
free backbone.  Finally, we would repeat the procedure on all
colorable 2-edge-removed remainder graphs, potentially eliminating
pairs from the frozen and free backbones each time.  Comparing with
the exact testbed arising from this procedure shows that, for all
graphs, EO found the correct ground state cost.  Moreover, EO
overestimated the frozen backbone fraction on only 4 graphs out of 440
(2 at $n=128$ and 2 at $n=256$, in both cases at $\alpha=4.6$), and by
at most 0.004.  This leads to a predicted systematic error that is at
least an order of magnitude smaller than the statistical error bars in
the results we present later.  EO's free backbone results were not of
as good quality, overestimating the backbone fraction on 36 graphs out
of 440, by an average of 0.003 though in one case ($n=256$,
$\alpha=4.6$) by as much as 0.027.  The resulting systematic error,
however, is still small compared to the statistical error bars in our
main results.

\begin{figure}
\vskip 2.4in
\includegraphics{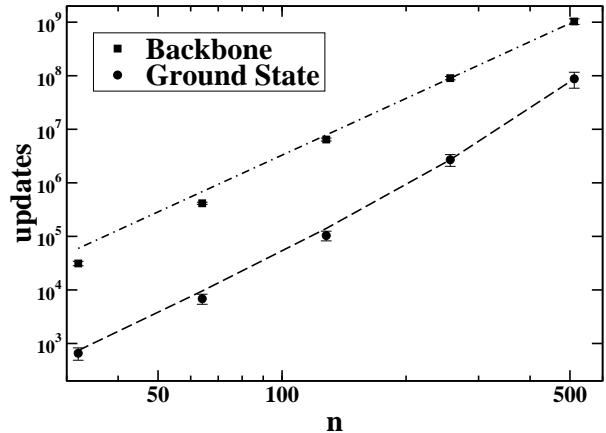}
\caption{Log-log plot of the average time $\left<t_{\rm GS}\right>$ to
reach the lowest cost state found (circles) and $\left<t_{\rm
BB}\right>$ to sample the backbone (squares), in units of EO-update
steps, as a function of system size $n$. The dashed line,
$0.003n^{3.5}\exp(0.004n)$ gives a reasonable fit to $\left<t_{\rm
GS}\right>$ (after having assumed the $n^{3.5}$ power law for the
fit), and the dash-dotted line, $0.3n^{3.5}$, is a fit obtained for
$\left<t_{\rm BB}\right>$. (Taking these crude fits at face value,
merely reaching the first good ground state approximations would
begin to dominate the runtime at about $n\approx10^3$).  }
\label{3coltiming}
\end{figure}

It is also instructive to study the running times for EO, how they
scale with increasing graph size, and how they compare with the exact
algorithm we have used for benchmarking.  In our EO implementation, we
have measured the average number of update steps it took (1) to find
the ground state cost for the first time, $\left<t_{\rm GS}\right>$,
and (2) to sample the backbone completely, $\left<t_{\rm BB}\right>$.
(Note that $t_{\rm BB}$, corresponding to $r_0$ in
Sec.~\ref{backbone}, is always less than half the total time spent to
satisfy the stopping criterion for an EO run described above.)  Both
$t_{\rm GS}$ and $t_{\rm BB}$ can fluctuate widely for graphs of a
given $n$ and $\alpha$, especially when $C>0$. However, since our
numerical experiments involve a large number of graphs, the average
times $\left<t_{\rm GS}\right>$ and $\left<t_{\rm BB}\right>$ are
reasonably stable. Furthermore, $\left<t_{\rm GS}\right>$ and
$\left<t_{\rm BB}\right>$ show only a weak dependence on $\alpha$,
varying by no more than a factor of 2: $\left<t_{\rm GS}\right>$
increases slowly for increasing $\alpha$, and $\left<t_{\rm
BB}\right>$ has a soft peak at $\alpha_{\rm c}$~\footnote{While search
times to solve the decision problems should be sharply peaked here,
determining the lowest cost solutions remains hard when $C>0$ for
$\alpha>\alpha_{\rm c}$, leading more to a plateau than to a peak. It
may also be the case that a search with EO is less influenced by the
transition, as other studies~\cite{eo_perc} have suggested, but the
range around $\alpha_{\rm c}$ that we have studied here is too small
to be conclusive on this question.}.  Thus, for each $n$ we average
these times over $\alpha$ as well, leading only to a slight increase
in the error bars (on a logarithmic scale).  We have plotted the
average quantities in Fig.~\ref{3coltiming} on a log-log scale as a
function of $n$. It suggests that the time EO takes to find ground
states increases exponentially but very weakly so.  Once they are
found, EO manages to sample a sufficient number of them in about
$O(n^{3.5})$ updates to measure the backbone accurately.

By contrast, one cannot easily quantify the scaling behavior of
running times for the exact branch-and-bound benchmarking method.  As
ground state cost increases, the complexity of the method quickly
becomes overwhelming, and rules out using it to measure average
quantities with any statistical significance. Clearly,
branch-and-bound itself has exponential complexity for determining
colorability. For the sizes studied here, however, the exponential
growth in $n$ appears in fact sub-dominant to the $O(n^{C+2})$
complexity of evaluating the backbone for a graph with non-zero ground
state costs.  When $C=1$ or $2$, the combinatorial effort is
manageable, but at $n=256$, graphs just at the transition
($\alpha=4.70$) reach $C\geq3$ and the algorithm takes weeks to test
all remainder graphs. From this comparison, one can appreciate EO's
speed in estimating the backbone fractions, however approximate!

\section{Numerical results}
\label{numerics}
\begin{figure}
\vskip 2.4in \includegraphics{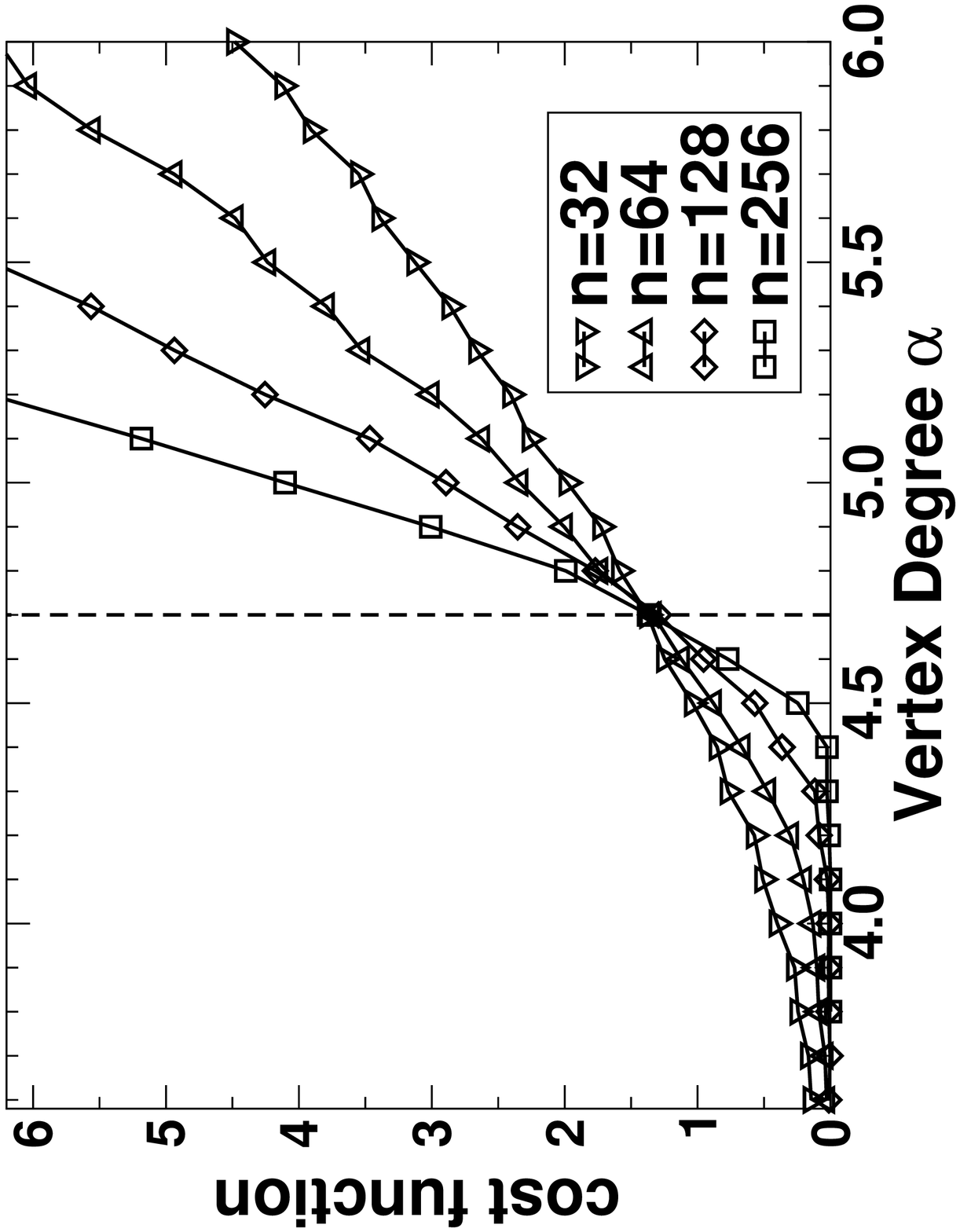} \includegraphics{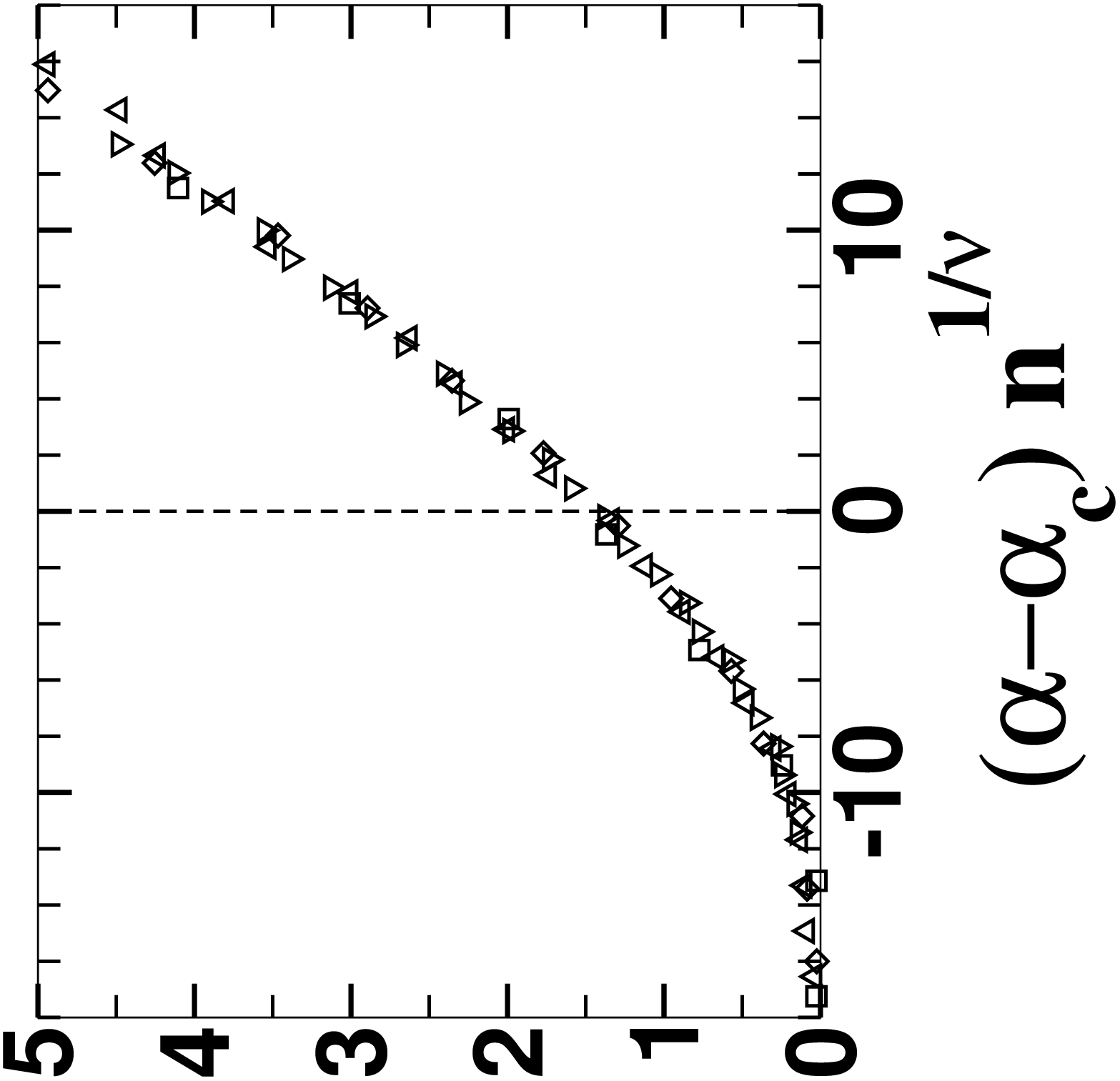}
\caption{ Plot of the average cost as a function of the vertex
degree $\alpha$. After correct finite size
scaling, the data collapses onto a single scaling function, as shown
in the insert. The fit gives $\alpha_{\rm crit}\approx4.70$, marked by
a vertical line.}
\label{3colplot}
\end{figure}

With the EO implementation as described above, we have sampled ground
state approximations for a large number of graphs at each size $n$. In
particular we have considered, over a range of $\alpha$, 100\,000
random graphs of size $n=32$, 10\,000 of size $n=64$, 4\,000 of size
$n=128$, and 1\,000 of size $n=256$.  By averaging over the lowest
energies found for these graphs, we obtain an approximation for
average ground state costs $\left<C\right>$ as a function of $\alpha$
and $n$, as shown in Fig.~\ref{3colplot}.  We have also sampled 160
instances of size $n=512$, which provided enough statistics for the
backbone though not for the ground state costs.

With a finite size scaling ansatz
\begin{eqnarray}
\left< C\right> \sim n^{\delta} f\left[\left(\alpha-\alpha_{\rm
crit}\right) n^{1/\nu}\right],
\label{scalingeq}
\end{eqnarray}
systematically applied~\cite{Somen}, it is possible to extract precise
estimates for the location of the transition $\alpha_c$ and the
scaling window exponent $\nu$.  In the scaling regime, one might
assume that the cost for the fixed argument of the scaling function is
independent of the size, i.e., $\delta=0$, indicated by the fact that
for all values of $n$ the cost functions cross in virtually the same
point.  Hence, in results we have previous reported~\cite{eo_prl}, we
obtained what appeared to be the best data collapse by fixing
$\delta=0$ and choosing $\alpha_{\rm crit}=4.72(1)$ and $\nu=1.53(5)$,
with the error bars in parentheses being estimates based on our own
assessment of the data collapse.  But a more careful automated fit to
our data, provided to us by S.M.~Bhattacharjee, gives
$\delta\approx-0.001(3)$, $\alpha_{\rm crit}=4.703(28)$, and
$\nu=1.43(6)$ with a tolerance level of $\eta=1\%$ (see
Ref.~\cite{Somen}). While these fits are consistent with our previous
results, they are also consistent with and much closer to the
presumably exact result of $\alpha_{\rm crit}=4.69\ldots$~\cite{MPWZ},
and the error estimates are considerably more trustworthy.

The scaling window is determined by two competing contributions: for
the intermediate values of $n$ accessible in this study it is
dominated by nontrivial contributions arising from the correlations
amongst the variables, which yields $\nu\approx1.43(6)$, similar to
satisfiability problems~\cite{Selman}. However, for sufficiently large
$n$, Wilson~\cite{Wilson} has shown that $\nu\geq2$, due to intrinsic
features of the ensemble of random graphs. The argument may be
paraphrased as follows. Since $\alpha=O(1)$ and vertex degrees are
Poisson-distributed with mean $\alpha$, a finite fraction of vertices
in a random graph have degrees 0, 1, or 2 (those not belonging to the
3-core~\cite{PSW}) and thus cannot possibly cause monochromatic
edges. But this finite fraction itself undergoes (normal) $\sim
1/\sqrt{n}$ fluctuations, and these fluctuations limit the
narrowing of the cost function's scaling window at large $n$. Such
variables make up about 15\% of the total near $\alpha_{\rm crit}$, so
we estimate the crossover to occure at $n^{-1/\nu}\sim0.15n^{-1/2}$ or
$n\approx10^5$, assuming all other constants to be unity.

\begin{figure}
\vskip 4.5in \includegraphics{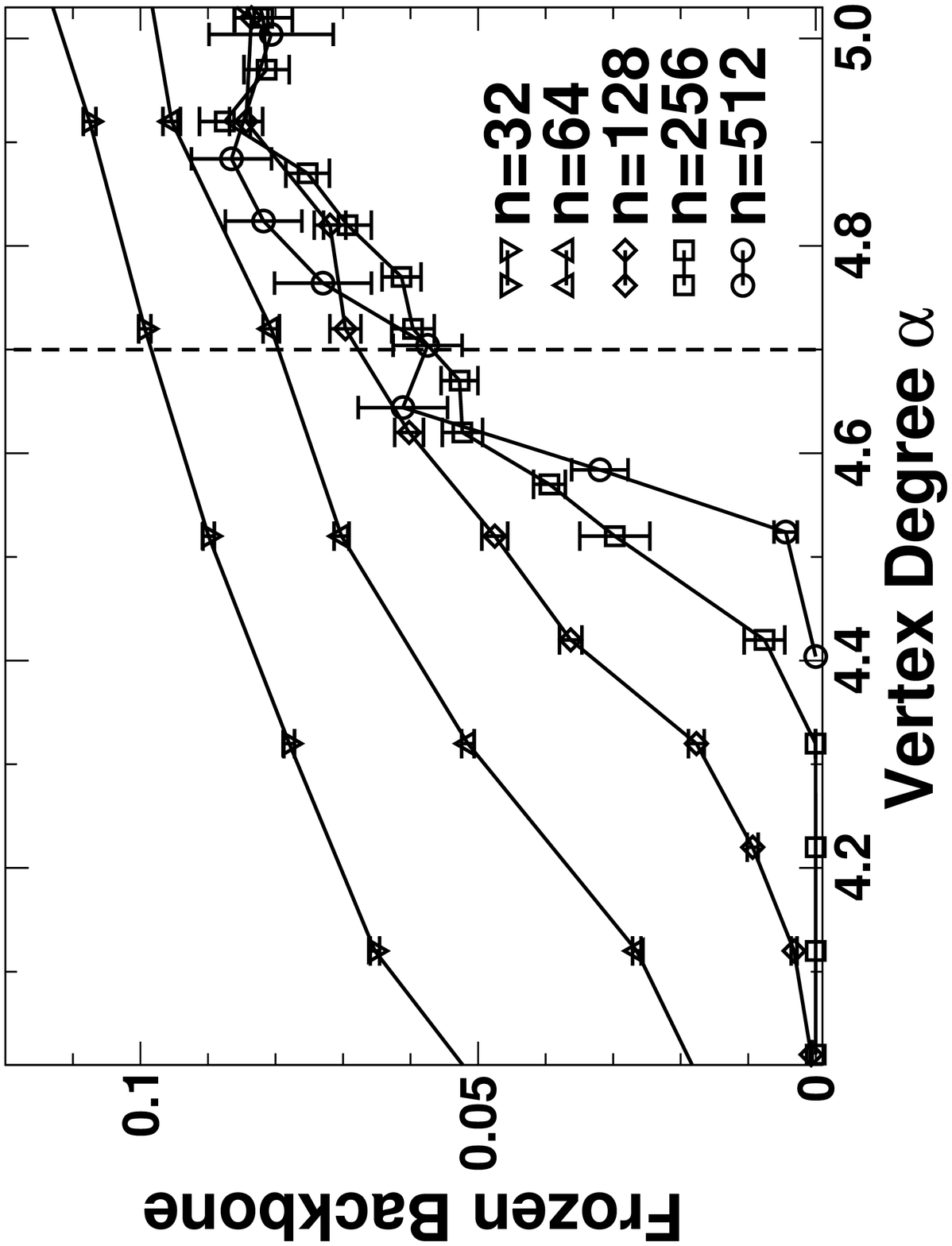} \includegraphics{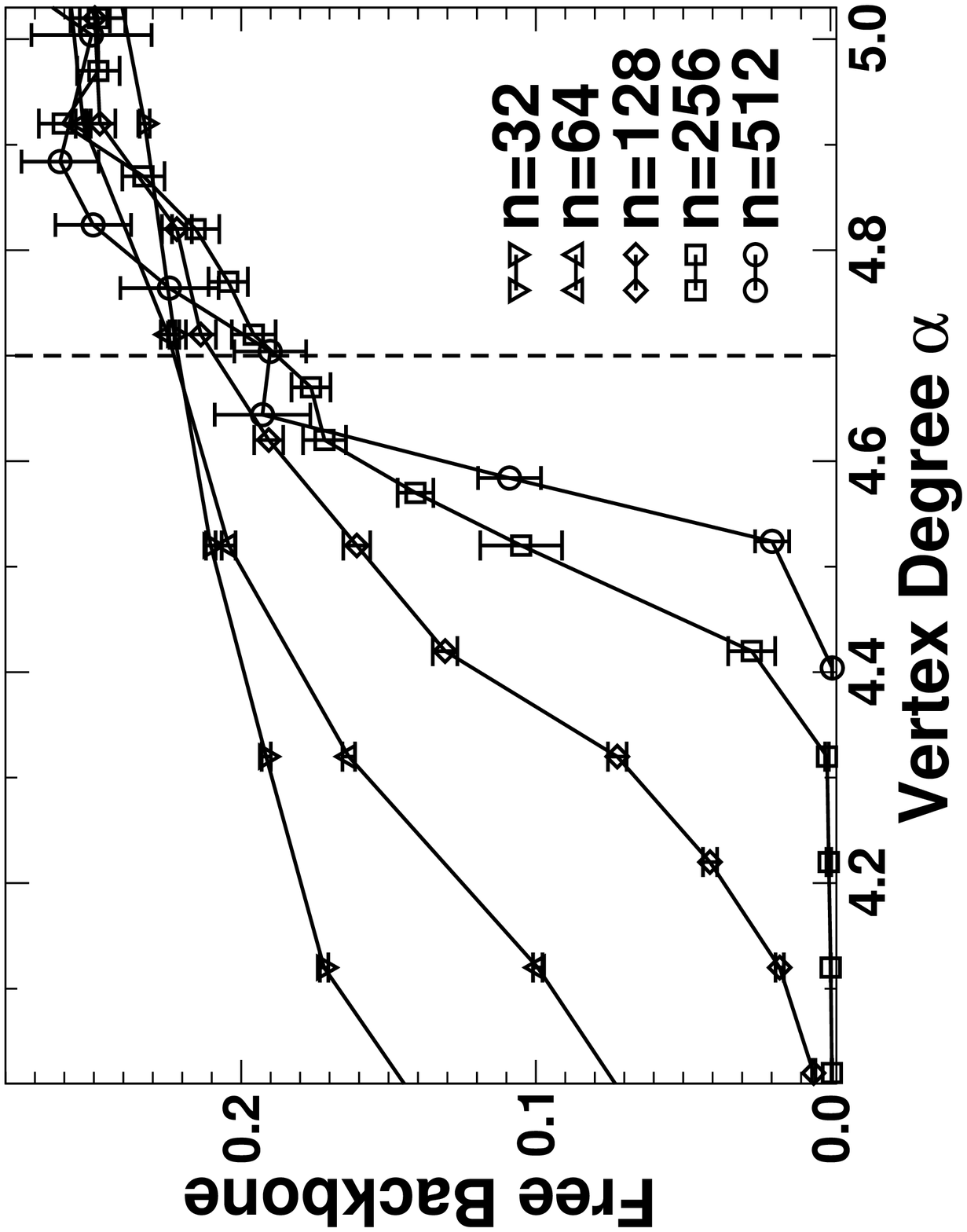}
\caption{ Plot of the frozen (top) and the free (bottom) backbone
fraction as a function of the vertex degree $\alpha$. The critical
point $\alpha_{\rm crit}\approx4.70$ is indicated by a vertical line.}
\label{backboneplot}
\end{figure}

Our next main result is the estimate of the backbone near the phase
transition, as described in Sec.~\ref{backbone}. We have sampled the
frozen and the free backbones~\cite{Culberson} separately. Our results
show the fraction of vertex pairs in each backbone, and are plotted in
Fig.~\ref{backboneplot}. For the free backbone, consistent with our
definition, we do not include any pairs that are already connected by
an edge. Although they make up only $O(1/n)$ of the pairs, the
inclusion of these would cause a significant finite size effect when
the backbone is small, and only by omitting them does the free
backbone vanish for $\alpha<\alpha_{\rm crit}$. In principle,
according to our definition one should also be sure to exempt from the
frozen backbone any pairs that are connected by a monochromatic edge
in all ground state configurations, but at $\alpha_{\rm crit}$ their
impact on the backbone is only $O(1/n^2)$.

As Fig.~\ref{backboneplot} shows, both backbones appear to evolve
toward a discontinuity for increasing $n$. The backbone fraction comes
increasingly close to vanishing below $\alpha_{\rm crit}$, followed by
an increasingly steep jump and then a plateau that, to within
statistical noise, appears stable at large $n$.  The height of the
plateau at $\alpha_{\rm crit}$ suggests that on average about 6\% of
all pairs are frozen and close to 20\% are free, with both values
rising further for increasing degree.  The ``jump'' in the frozen
backbone is somewhat smaller than that in the free backbone, adding a
higher degree of uncertainty to that interpretation, although still
well justified within the error bars. Indeed, given the considerable
ground state degeneracies, it would be surprising if the frozen
backbone were large.

\begin{figure}
\vskip 4.6in 
\includegraphics{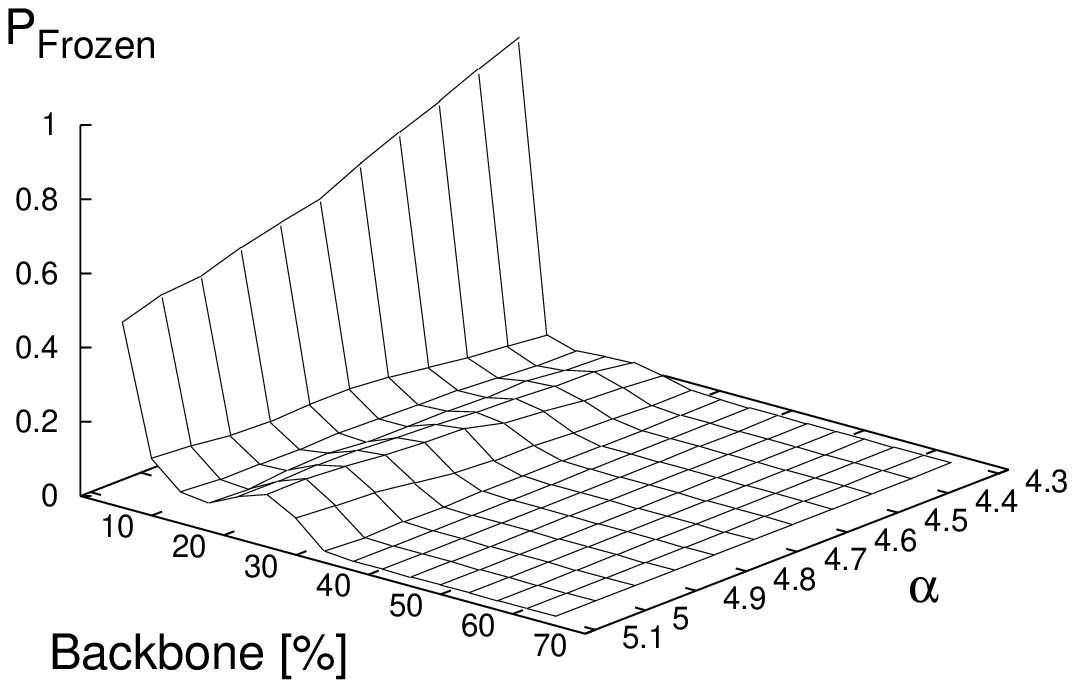} 
\includegraphics{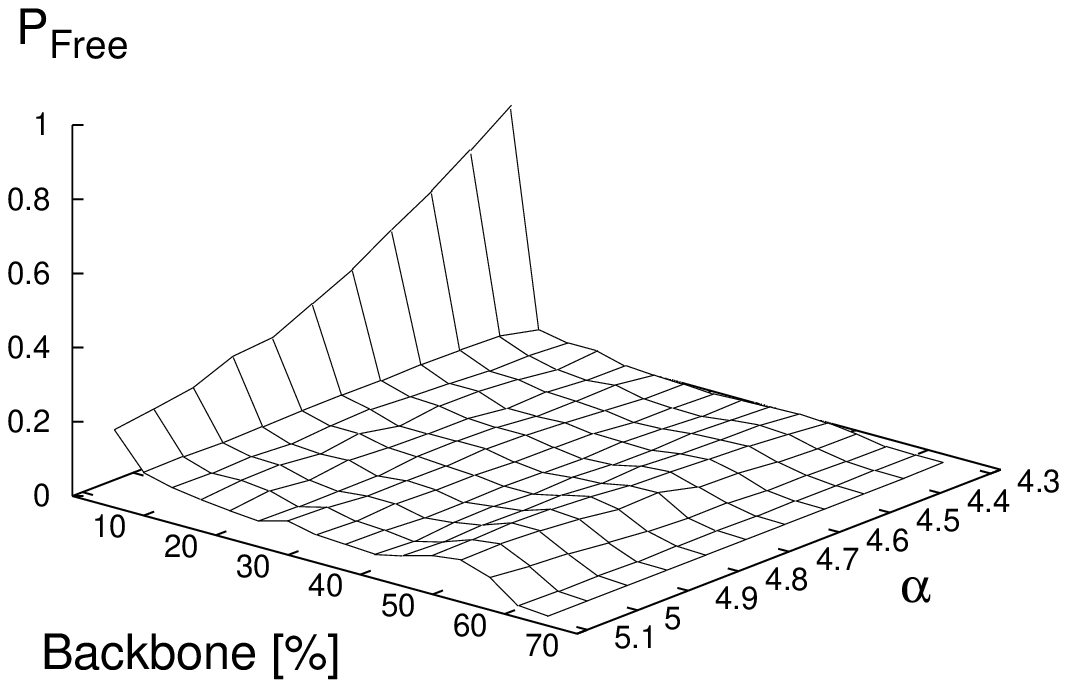}
\caption{Plot of the typical frozen (top) and free (bottom) backbone
probability, obtained here from the $n=128$ graphs. For mean degree
values $\alpha<\alpha_{\rm crit}$, graphs are almost certain to have a
vanishing backbone, both frozen and free. Above $\alpha_{\rm crit}$ a
majority of graphs still do not exhibit any backbone, but a finite
fraction of graphs display a sizable backbone fraction, clustered at a
characteristic size. The average backbones plotted in
Figs.~\protect\ref{backboneplot} represent the average of these
apparently bimodal distributions. This qualitative picture appears to
hold for increasing $n$, although $n>128$ data are somewhat
noisy. }
\label{bbprob}
\end{figure}

A more detailed look at the data (Fig.~\ref{bbprob}) suggests that the
distribution of frozen backbone fractions for individual instances is
bimodal at the transition, i.e., about half of the graphs have a
backbone well over 10\% while the other half have no backbone at all,
leading to the average of 6\% mentioned above. Furthermore, there
appears to be some interesting structure in the backbone
discontinuity, which may be significant beyond the noise. Note in
Fig.~\ref{backboneplot} that for larger $n$, the increase of the
frozen backbone stalls or even reverses right after the jump before
rising further. This property coincides with the emergence of non-zero
costs in the ground state colorings (see Fig.~\ref{3colplot}). The
sudden appearance of monochromatic edges seems initially to reduce the
frozen backbone fraction: typically there are numerous ways of placing
those few edges, often affecting the most constrained variables pairs
and eliminating them from the frozen backbone.  Similar observations
have been made by Culberson~\cite{Culberson}.  According to this
argument, only the frozen backbone should exhibit such a stall (or
dip). Indeed, Fig.~\ref{backboneplot} indeed shows a less hindered
increase in the free backbone, though the difference there may be
purely due to statistical noise.

\section{Conclusions}
\label{conclusion}
We have considered the phase transition of the MAX-3-COL problem for a
large number of instances of random graphs, of sizes up to $n=512$ and
over a range of mean degree values $\alpha$ near the critical
threshold.  For each instance, we have determined the fraction of
vertex pairs in the frozen and free backbones, using an optimization
heuristic called {\em extremal optimization} (EO)~\cite{BoPe1}. Based
on previous studies~\cite{eo_perc}, EO is expected to yield an
excellent approximation for the cost and the backbone.  Comparisons
with a testbed of exactly-solved instances suggest that EO's
systematic error is negligible compared to the statistical error.

Using a systematic procedure for optimizing the data collapse in
finite size scaling~\cite{Somen}, we have argued that the transition
occurs at $\alpha_{\rm crit}=4.703(28)$, consistent with earlier
results~\cite{Cheese,Hogg,Culberson,eo_prl} as well as with a recent
replica symmetry breaking calculation yielding 4.69~\cite{MPWZ}.  We
have also studied both free and frozen backbone fractions around the
critical region.  A simple argument~\cite{Monasson} demonstrates that
below the critical point the backbone fraction always vanishes for
large $n$.  At and above the critical point, neither backbone appears
to vanish, suggesting a first-order phase transition.  This is in
close resemblance to $K$-SAT for $K=3$~\cite{Monasson}; indeed, both
are computationally hard at the threshold.

Even though the backbone is defined in terms of minimum-cost
solutions, its behavior appears to correlate more closely with the
complexity of finding a zero-cost solution (solving the associated
decision problem) at the threshold.  One possible explanation is that
instances there have low cost, so finding the minimal cost is only
polynomially more difficult than determining whether a zero-cost
solution exists.  Interestingly, our 3-coloring backbone results
mirror those found for the spine~\cite{Culberson}, an upper bound on
the backbone that is defined purely with respect to zero-cost graphs.
The
authors of that study speculate that at the threshold, although the
spine is discontinuous, the backbone itself might be {\em continuous}.
Our results contradict this speculation, instead providing support for
a relation --- albeit restricted --- between backbone behavior and
average-case complexity.

\section*{Acknowledgments}
We are greatly indebted to Somen Bhattacharjee for providing an
automated evaluation of our finite size scaling fit. We wish to thank
Gabriel Istrate, Michelangelo Grigni, and Joe Culberson for helpful
discussions, and the Los Alamos National Laboratory LDRD program for
support.

\end{document}